\documentstyle[graphicx]{article}
\title{
NONDIFFRACTING OPTICAL BEAMS: \\
physical properties, experiments and
applications}
\author{Zden\v{e}k Bouchal \\
Department of Optics, Palack\'y University,\\
 17. listopadu 50,
772 07 Olomouc, \\
Czech Republic}
\date{}
\begin{document}
\maketitle
%%%%%%%%%%%%%%%%%
\begin{abstract}
The controversial term "nondiffracting beam" was introduced into
optics by Durnin in 1987.
Discussions related to that term revived interest in problems
of the light diffraction and resulted in an appearance of the new
research direction of the classical optics dealing with the
localized transfer of electromagnetic energy.
In the paper, the physical concept
of the  nondiffracting propagation is presented
and the basic properties of the nondiffracting beams
are reviewed. Attention is also focused to the experimental
realization and to applications of the nondiffracting beams.
\end{abstract}
%%%%%%%%%%%%%%%%%%%%%%%%%%%%%%%%%%%%%%%%%%%%%%%%%%%%%%%%%%%%
\section{INTRODUCTION}
~

Many phenomena observable in our everyday life indicate that light
propagates rectilinearly. Rectilinear propagation is one of the most
apparent properties of light. It serves as an argument that light is
a stream of particles. However, some optical phenomena and
experiments indicate that the law of rectilinear propagation of light
does not hold. They can be satisfactorily explained only on the assumption
that light is a wave. Historically, the diffraction effects are associated
with violation of the rectilinear propagation of light. By Sommerfeld,
diffraction is defined as any deviation of light from rectilinear
propagation, not caused by reflection or refraction. The strong diffraction
effects appear if the transverse dimensions of the beam of light are
comparable to the wavelength. The diffraction phenomena are
best appreciable for long waves such as sound or water waves.
In optics, the diffraction effects are less apparent. They are responsible
for the beam divergence in the free propagation and for penetration
of light into the region of the geometric shadow.
In the modern treatment, diffraction effects are not connected with
light transmission through apertures and obstacles only.
Diffraction is examined as a natural
property of wavefield with the nonhomogeneous transverse intensity
distribution. It commonly appears even if the beam is transversally
unbounded. The Gaussian beam is the best known example.

In optics, the nondiffracting propagation of the beam-like fields
can be obtained in convenient media such as waveguide or nonlinear
materials. The beams then propagate as waveguide modes and spatial
solitons, respectively. In 1987, the term nondiffracting beam
appeared also in relation to the propagation in vacuo
\cite{durnin}. The nondiffracting beam was comprehended as the
monochromatic optical field whose transverse intensity profile
remains unchanged in free-space propagation. In the original
Durnin's paper, the beams were examined as exact solutions to the
homogeneous Helmholtz equation. They were obtained in the system of the
cylindrical coordinates under restriction that their complex
amplitude is separable as the product of the functions $R(r)$,
$\Phi(\varphi)$ and $Z(z)$ depending on the coordinates
$r,\varphi$ and $z$, respectively.
The transverse amplitude profile of such beams can be described
by the Bessel functions so that they are usually called Bessel beams.
Later, the more general types of nondiffracting beams were
introduced \cite{bouchal1,rushin,new} and the properties by which
they become different from the common laser beams were examined.
Recently, the method enabling generation with the transverse
intensity profile which can be predetermined and controlled has
been proposed and examined \cite{bouchol}.
The particular attention was focused on the analysis of the
admissible amplitude profiles of the nondiffracting beams and on
their wavefront properties. The nondiffracting beams originally
analysed in the scalar approximation were generalized to the
vectorial electromagnetic beams exactly fulfilling the Maxwell
equations \cite{mishra,tufr,bouchal2,bbb}.
Propagation invariance of the intensity profile of the
nondiffracting beams was explained as a result of the
convenient composition of the angular spectrum. It composes
the plane waves whose propagation vectors are placed on the conical
surface. Mathematically, such angular spectrum can be described
by the Dirac delta function $\delta(\nu-\nu_{0})$, where
$\nu_{0}$ is the single radial spatial frequency representing
the basic beam parameter.
The ideal nondiffracting beam then arises as an
interference field produced by the coherent superposition of the
plane waves whose relative phase differences remain unchanged in
the free propagation.
By that way, the diffraction effects can be overcome
in free propagation of the
source-free monochromatic wavefields or pulses.
However, the ideal nondiffracting beams possessing the sharp
$\delta$-like angular spectrum  carry an infinite energy.
That is a reason, why the
diffraction cannot be overcome in real situations, and why the
nondiffracting beams cannot be exactly realized. In experiments,
only approximations known as the pseudo-nondiffracting
beams can be obtained \cite{gori,overfelt}.
An idea about properties of the realizable beams with the finite
energy can be simply obtained if the ideal nondiffracting beam is
bounded by the homogeneously transmitting aperture of finite
dimensions or by the Gaussian aperture. The propagation
invariance of the transverse intensity profile of the
nondiffracting beam impinging on the aperture is lost and the
beam behind the aperture propagates with the diffractive
divergence.
Regardless of that fact, there are
important distinctions between propagation properties of the
pseudo-nondiffracting and the conventional, for example Gaussian
beams. They are usually demonstrated in numerical simulations.
Recently, the distinct diffractive divergences of the
pseudo-nondiffracting and the conventional beams have been
explained and interpreted by means of the uncertainty relations
and demonstrated experimentally \cite{bouchal3}.

Except of the fully eliminated diffractive divergence of the
ideal nondiffracting beams and the reduced diffractive spread
of the pseudo-nondiffracting beams their further peculiar
properties useful for applications were explored.
Attention was focused on the robustness of the beams
manifested by their resistance against amplitude and phase
distortions.
It was shown that the ideal nondiffracting beam disturbed by a
nontransparent obstacle is able to regenerate its intensity
profile to the original form in the free propagation behind the
obstacle \cite{mc}. In \cite{chlup}, the effect was described for both the
nondiffracting and the pseudo-nondiffracting beams
and verified by the simple experiment.

The coherent superposition of the nondiffracting modes resulting
in the self-imaging effect was examined in \cite{ojeda,bouchal4}.
The effect appears if the angular wavenumbers of the
nondiffracting modes are conveniently coupled. It represents
the spatial analogy with the mode-locking realized in the
temporal domain. Due to the interference of the modes, the
transverse intensity profile of the beam reappears periodically
at the planes of the constructive interference and vanishes
at the planes of the destructive interference.

Recently, the self-imaging effect was used for periodical
self-reconstruction of the coherent light field with an arbitrary
predetermined amplitude profile.
Theoretical description of the
effect was proposed for both the monochromatic and the nonstationary
wavefields \cite{bouchal5,bhorak,bert}
and the experimental verification was realized
applying the special Fourier filter used in the 4-f optical
system \cite{wagner}.
The controllable 3D spatial shaping of the
coherent optical fields has also been proposed and examined.
That methods enables localization of the light energy
into the small volume elements with the size comparable to
the wavelength \cite{boukyvn}.

During last decade an increasing attention has been given to the
wavefields possessing the line, spiral or combined wavefront
dislocations. In optics, such fields are known as the optical
vortices \cite{ind,bas}.
Some types of nondiffracting beams can also belong
to the class of optical vortices. The vortex beam is
characterized by the topological charge and its phase
singularities and the helical wavefront can be visualized by the
interferometric methods. The wavefront helicity of the vortex
beam is associated with the spiral flow of the electromagnetic
energy. That property was successfully applied in
experiments testing the transfer of the angular momentum of the
electromagnetic field to the microparticles \cite{power}.
Recently, the effect of the self-regeneration of the
nondiffracting vortex beam appearing after interaction with
microparticles has been verified \cite{bn1,bn2}.

The basic concept of the nondiffracting propagation has been
developed for the fully coherent light.
Recently, attention has been focused also
to an interesting task to join the problems of the
variable-coherence optics with the nondiffracting
propagation of light beams.
A general description of the partially coherent
propagation-invariant fields including the partially coherent
nondiffracting beams has been proposed in
\cite{turunen,friberg}.
Directionality of the partially-coherent Bessel-Gauss beams
has been analysed on the assumption that the beams are produced
from a globally incoherent source \cite{zahid}.
Propagation properties of the partially coherent
pseudo-nondiffracting beams obtained by
the incoherent superposition of
identical coherent beams whose propagation axes lie on a conical
surface have been examined in \cite{wolf}.
Recently, the nondiffracting beams with the controlled spatial
coherence have been introduced and analyzed \cite{bouchal6}.

The ideal nondiffracting beams can be obtained
as a superposition of the plane waves whose radial angular frequencies
are restricted by the Dirac delta function to the single value.
In the geometrical interpretation,  the propagation vectors of the plane wave
components of the angular spectrum form the conical surface.
In optics, several experiments have been proposed to
produce a good approximation of the required
composition of the angular spectrum.
The original Durnin's experimental demonstration of the
zero-order Bessel beam utilized an annular slit placed at the
focal plane of the lens \cite{miceli}.
More efficient methods of generating
the required conic wavefront based on the use of the
computer-generated holograms \cite{vasara}, the axicon
\cite{indeb,scott} or the programmable spatial light
modulators \cite{davis} were also suggested.
Experimental realization of the nondiffracting Bessel
beam due to the spherical aberration of the simple lens
\cite{herman} and by means of the two-element refracting system
\cite{karim} has been successfully performed.
Experimental methods applicable to generation of the
pseudo-nondiffracting beams were reviewed in \cite{lapointe}.
Their applications have been proposed in the field of acoustics,
metrology and nonlinear optics \cite{lu,arimoto,wulle}.
Properties of the nondiffracting beams are perspective for the design
of the electron accelerators \cite{romea} and the optical
tweezers \cite{dholakia}.
%%%%%%%%%%%%%%%%%%%%%%%%%%%%%%%
\section{COHERENT NONDIFFRACTING BEAMS}
\subsection{Concepts of nondiffracting propagation}
~

The ideal monochromatic spatially coherent nondiffracting
beam propagating along the z-axis is comprehended as
the mode-like field whose complex amplitude can be written
in the form
%%%%%%%%%%%%%%%%%%%%%%%%%%%%%%%
\begin{eqnarray}
U(x,y,z,t)=u(x,y)\exp[i(\omega t-\beta z)],\label{def}
\end{eqnarray}
%%%%%%%%%%%%%%%%%%%%%%%%%%%%%%%
where $u$ describes the transverse amplitude profile and
$\omega$ and $\beta$ are the angular frequency
and the angular wavenumber, respectively.
The slowly varying amplitude $u$ is then independent of the
$z$-coordinate so that the intensity of the beam
$I=UU^{*}$ is propagation invariant. The fields (\ref{def})
are known as waveguide modes or spatial solitons propagating
in optical linear and nonlinear materials but
Durnin's original work \cite{durnin} has excited interest also
in their free-space propagation. In that case, the complex
amplitudes $U$ must fulfil the homogeneous wave equation.
The temporally independent amplitude
%%%%%%%%%%%%%%%
\begin{eqnarray}
a(x,y,z)=u(x,y)\exp(-i\beta z)\label{sep}
\end{eqnarray}
%%%%%%%%%%%%%%%
then fulfils the Helmholtz equation
%%%%%%%%%%%%%%%%%%%%%%%
\begin{eqnarray}
\left(\nabla^{2}+k^{2}\right)a(x,y,z)=0,\label{he}
\end{eqnarray}
%%%%%%%%%%%%%%%%%%%%%%%
where $k=\omega/c$ and $c$ is the light velocity in vacuo.
The mathematical
description of the ideal monochromatic nondiffracting beam
can be based on the differential or on the integral formalism.
%%%%%%%%%%%%%%%%%%%%%%%%%%%%%%%
\subsubsection{Separable solutions to the Helmholtz equation}
~

The spatial evolution of the complex amplitude $a$
can be described by the transverse and the longitudinal parts
depending only on the transverse coordinates $(x,y)$ and on the
$z$-coordinate, respectively. Though the homogeneous
(source-free) Helmholtz equation can be separated in 11
coordinate systems, the required separability into
the transverse and the longitudinal parts is possible only in
Cartesian, circular cylindrical, parabolic cylindrical,
and elliptical cylindrical coordinates. The particular attention
has been  focused on the circular cylindrical and
the elliptical cylindrical
coordinates for which the transverse amplitude profile
$u$ can be expressed by the known functions.\\[10pt]
%%%%%%%%%%%%%%%%%%%%%%%%%%%%%%%
{\it Circular cylindrical coordinates}
~

The circular cylindrical coordinates  $(r,\varphi ,z)$ are
related to the Cartesian coordinates $(x,y,z)$ by
$x=r\cos{\varphi}$, $y=r\sin{\varphi}$ and $z=z$
where $r\in <0,\infty)$ and $\varphi\in <0,2\pi>$.
The solutions of the Helmholtz equation (\ref{he})
then can be found only under restricting assumption
that the amplitude $u$ can be expressed as a product
of the functions $R$ and $\Phi$ depending on the
radial coordinates $r$ and $\varphi$, respectively.
In that case, the complex amplitudes $a$ are assumed to be of the
form
%%%%%%%%%%%%%%%%%%%%%%%%%%%%%
\begin{eqnarray}
a(r,\varphi,z)=R(r)\Phi(\varphi)\exp(-i\beta z).\label{amp1}
\end{eqnarray}
%%%%%%%%%%%%%%%%%%%%%%%%%%%%%
The function $\Phi$ describing dependence of the transverse
amplitude profile of the beam must be periodical. Usually
we assume that it is of the form
%%%%%%%%%%%%%%%%%%%%%%%%%%%%%
\begin{eqnarray}
\Phi(\varphi)=\exp(im\varphi),\ \ m=0,1,2,\cdots.\label{amp2}
\end{eqnarray}
%%%%%%%%%%%%%%%%%%%%%%%%%%%%%
Substituting (\ref{amp1}) and (\ref{amp2}) into the Helmholtz
equation (\ref{he}) we obtain the differential equation
for the radial function $R$. It is known as the
Bessel equation
%%%%%%%%%%%%%%%%%%%%%%%%%%%
\begin{eqnarray}
\frac{d^{2}R(r)}{dr^{2}}+\frac{1}{r}\frac{dR(r)}{dr}
+\alpha^{2}R(r)\left(1-\frac{m^{2}}{\alpha^{2}r^{2}}\right)
=0, \label{be}
\end{eqnarray}
%%%%%%%%%%%%%%%%%%%%%%%%%%%
$\alpha^{2}=k^{2}-\beta^{2}$.
Its general solution can be given as a linear combination
of the $m$-th order Bessel functions of the first kind
$J_{m}$ and the $m$-th order Neumann functions $N_{m}$
\cite{arfken},
%%%%%%%%%%%%%%%%%%%%%%%
\begin{eqnarray}
R_{m}(r)=\mu J_{m}(\alpha r)+\nu N_{m}(\alpha r),
\end{eqnarray}
%%%%%%%%%%%%%%%%%%%%%%%
where $\mu$ and $\nu$ are the weighing coefficients.
Usually, the Bessel functions of the first
kind are considered as the only physical solutions.
The reason why the Neumann functions are considered as the
unphysical solutions is that they possess singularities
at the zero point if they are used separately.
However, as has been shown in \cite{new},
in combination with the Bessel
functions of the first kind, the Neumann functions have a
physical meaning. Two linear combinations yielding the $m$-th
order Hankel functions are of particular importance.
The radial functions obtained as the solutions of the Bessel
equation can be denoted as $R^{j}\equiv H^{j}_{m}$, $j=1,2$,
where the Hankel functions $H^{j}_{m}$ are given as
%%%%%%%%%%%%%%%%%
\begin{eqnarray}
H_{m}^{1}&=&J_{m}(\alpha r)+iN_{m}(\alpha r),\label{h1}\\
H_{m}^{2}&=&J_{m}(\alpha r)-iN_{m}(\alpha r).\label{h2}
\end{eqnarray}
%%%%%%%%%%%%%%%%%
They have simple interpretation for the zero-order Hankel
function \cite{new}.
%To demonstrate it, we consider the spatial
%distribution of the beam amplitude for the time $t$ such that
%$\exp(i\omega t)=1$.
The real amplitude of the rotationally symmetrical beam
can be defined as
%%%%%%%%%%%%%%%%%%%
\begin{eqnarray}
\overline{U}(r,z)=\frac{1}{2}\left [U(r,z)+U^{*}(r,z)\right ],
\end{eqnarray}
%%%%%%%%%%%%%%%%%%%
where
%%%%%%%%%%%%%%%%%%%%
\begin{eqnarray}
U(r,z)=\frac{1}{2}\left[H_{0}^{1}(\alpha r)
+H_{0}^{2}(\alpha r)\right ]\exp[i(\omega t-\beta z)].
\end{eqnarray}
%%%%%%%%%%%%%%%%%%%%%
Applying (\ref{h1}) and (\ref{h2}),
the amplitude $\overline{U}$ can be rewritten to the form
%%%%%%%%%%%%%%%%
\begin{eqnarray}
\overline{U}(r,z)=\frac{1}{2}\left [\overline{U}_{0}^{1}(r,z)+
\overline{U}_{0}^{2}(r,z)\right ],
\end{eqnarray}
%%%%%%%%%%%%%%%%
where
%%%%%%%%%%%%%%%%%%%%
\begin{eqnarray}
\overline{U}_{0}^{1}(r,z)&=&J_{0}(\alpha r)\cos(\omega t-\beta z)
+N_{0}(\alpha r)\sin(\omega t-\beta z),\\
\overline{U}_{0}^{2}(r,z)&=&J_{0}(\alpha r)\cos(\omega t-\beta z)
-N_{0}(\alpha r)\sin(\omega t-\beta z).
\end{eqnarray}
%%%%%%%%%%%%%%%%%%%%
The real amplitude of the beam $\overline{U}$ is now expressed as
a superposition of two travelling waves given by the Hankel
functions.
The amplitude $\overline{U}_{0}^{1}$ describes the
rotationally symmetric outgoing wave travelling away from the
axis (Fig. 1a) while the amplitude $\overline{U}_{0}^{2}$
represents the incoming wave travelling towards the axis (Fig. 1b).
In Fig. 1 the waves are illustrated for the time $t$ such
that $\exp(i\omega t)=1$.
The singularity of the included Neumann function can be
interpreted as arising from the collapse of the incoming
cylindrical wave onto the beam axis which serves,
simultaneously, as the source from which the outgoing wave
emanates. To satisfy boundary conditions at $r=0$, each
cylindrical wave must be the complex conjugate of the other.
In that case, the imaginary parts of the Hankel functions
composing the Neumann functions cancel out so that the total
field is the standing wave represented by the Bessel function
(Fig. 1c).
\\[10pt]
%%%%%%%%%%%%%%%%%%%%%%%%%%%%%%%
{\it Elliptical cylindrical coordinates}
~

The system of elliptical cylindrical coordinates
$(\zeta,\eta,z)$ can be defined as
$x=h\cosh{\zeta} \cos{\eta}$, $y=h\sinh{\zeta} \sin{\eta}$
and $z=z$, where $\zeta\in <0,\infty)$ and $\eta\in <0,2\pi)$,
and $2h$ represents the distance between the foci
of an ellipse placed at the plane $(x,y)$ of the Cartesian
coordinate system \cite{grad}. Taking into account the
separation of the complex amplitude (\ref{sep})
and applying the elliptical cylindrical coordinates the
Helmholtz equation (\ref{he}) can be rewritten to the form
\cite{straton}
%%%%%%%%%%%%%%%%%%%%%%%%%%%%%%%%
\begin{eqnarray}
\frac{\partial^{2}u(\zeta,\eta)}{\partial \zeta^{2}}+
\frac{\partial^{2}u(\zeta,\eta)}{\partial \eta^{2}}+
\frac{\alpha^{2}h^{2}}{2}(\cosh{2\zeta}-\cos{2\eta})
u(\zeta,\eta)=0.\label{hem}
\end{eqnarray}
%%%%%%%%%%%%%%%%%%%%%%%%%%%%%%%
On the assumption that the complex amplitude $u$ can be written
as the product of the functions depending only on the variables
$\zeta$ and $\eta$, respectively, the Helmholtz equation
(\ref{hem}) can be split into the Mathieu differential equations.
Application of their solutions to the description of the optical
beams has been proposed in \cite{gouesbet}. In \cite{vega}
the propagation properties of the zero-order Mathieu beam
have been examined. Its complex amplitude can be written in
the form
%%%%%%%%%%%%%%%%%%%%%%%%%%%%%%%%%
\begin{eqnarray}
U(\zeta, \eta,z,t;q)=U_{0}Ce_{0}(\zeta;q)ce_{0}(\eta;q)
\exp[i(\omega t-\beta z)],
\end{eqnarray}
%%%%%%%%%%%%%%%%%%%%%%%%%%%%%%%%%
where the parameter $q=\alpha^{2} h^{2}/4$ carries information
about the radial spatial frequency $\alpha/k$
influencing the beam transverse size and the ellipticity
of the coordinate system $h$.
The beam possesses the highly localized intensity
distribution along the $x$-direction and the sharply peaked
quasi-periodic structure along the $y$-direction.
The simple experimental set-up providing a good approximation
of the zero-order Mathieu beam was proposed in \cite{vega1}.

The separable solutions of the Helmholtz equation obtained
applying the circular cylindrical and the elliptical cylindrical
coordinates represent only special examples of the nondiffracting
fields whose transverse intensity profiles can be described by
the known functions. The more general nondiffracting fields can
be effectively examined applying the integral formalism.
%%%%%%%%%%%%%%%%%%%%%%%%%%%%%%%%%
\subsubsection{Integral form of the nondiffracting beam}
~

In the integral form, the temporally independent amplitude of
the general nondiffracting beam can be conveniently expressed
applying the circular cylindrical coordinates ${\bf r}\equiv
(r,\varphi,z)$
%%%%%%%%%%%%%%%%%%%%%%%%%%%%%%%
\begin{eqnarray}
a({\bf r})=\frac{ik}{2\pi}
\int_{-\pi}^{\pi}
A(\psi)f({\bf r},\psi)d\psi,\label{integ}
\end{eqnarray}
%%%%%%%%%%%%%%%%%%%%%%%%%%%%%%%
where
%%%%%%%%%%%%%%%%%%%%%%%%%%%%%%
\begin{eqnarray*}
f({\bf r},\psi)=\exp(-i\beta z)\exp[i\alpha r\cos(\psi-\varphi)],
\end{eqnarray*}
%%%%%%%%%%%%%%%%%%%%%%%%%%%%%%
and $A$ denotes an arbitrary periodical function.
The parameter $\alpha$ can be interpreted by means of the
angular spectrum. The angular spectrum $F$ is a function of the
angular frequencies $\nu_{x}$ and $\nu_{y}$ and can be obtained
as the two-dimensional Fourier transformation of the amplitude $a$.
Applying the radial angular frequency $\nu$ defined as
$\nu_{x}=\nu\cos{\psi}$ and $\nu_{y}=\nu\sin{\psi}$
it can be expressed as
%%%%%%%%%%%%%%%%%%%%%%%%%%%%%%%
\begin{eqnarray}
F(\nu,\psi)=A(\psi)\delta(\nu-\nu_{0}),\label{cas}
\end{eqnarray}
%%%%%%%%%%%%%%%%%%%%%%%%%%%%%%%
where
%%%%%%%%%%%%%%%%%%%%%%%%%%%%%%%
\begin{eqnarray}
\nu_{0}=\frac{\alpha}{2\pi}.\label{radfr}
\end{eqnarray}
%%%%%%%%%%%%%%%%%%%%%%%%%%%%%%%
The peculiar propagation properties of the ideal nondiffracting
beams appear as a consequence of the composition of the angular
spectrum. It contains only the single radial frequency $\nu_{0}$
so that the relative phases of the plane wave components remain
unchanged under propagation. In geometrical interpretation
the angular spectrum represents a coherent superposition
of plane waves whose propagation vectors cover the conical surface
with the vertex angle $2\theta_{0}=2\arcsin(\lambda \nu_{0})$,
where $\lambda $ denotes the wavelength. The amplitudes and
relative phases of the superposed plane waves can be arbitrary.
They are described by the function $A$. Due to
that arbitrariness an
infinite number of the nondiffracting beams with different
transverse intensity profiles can be obtained.
The parameters $\alpha$ and $\beta$ of the nondiffracting beam
have a simple geometrical interpretation. They represent
projections of the propagation vectors of the plane wave components
of the angular spectrum to the transverse plane $(x,y)$ and
to the direction of propagation coinciding with the $z-$axis,
respectively. They can be expressed as
$\alpha=k\sin{\theta_{0}}$ and
$\beta=k\cos{\theta_{0}}$.
%%%%%%%%%%%%%%%%
\subsection{Types of coherent nondiffracting beams}
~

The coherent nondiffracting field can be comprehended
as the interference field produced by the interference of
plane waves whose propagation vectors create the conical surface.
Due to the interference the total field can possess an
appreciable beam-like intensity peak so that it is often called
the nondiffracting beam. The condition (\ref{cas}) laid on the
composition of its angular spectrum
represents necessary and sufficient condition
of the nondiffracting propagation. The various intensity
profiles of the nondiffracting fields can be obtained by
manipulations of the amplitudes and phases of the plane
wave components of the angular spectrum. Because an arbitrary
azimuthal modulation of the angular spectrum is admitted
there exists an infinite number of nondiffracting fields.
Here, only best known examples are reviewed.
%%%%%%%%%%%%%%%%%%%%%%%
\subsubsection{Bessel beams}
~

In the special case when the amplitudes of the plane wave
components of the angular spectrum are constant and their
phases are azimuthally modulated by $A(\psi)=A_{0}\exp(im\psi)$
the integral representation (\ref{integ}) results in the
Bessel nondiffracting beams. Their
complex amplitude can be written in the form
%%%%%%%%%%%%%%%%%%%%%%%%%%%%%%%
\begin{eqnarray}
U(r,\varphi,z,t)=A_{0} J_{m}(\alpha r)
\exp[i(\omega t+m\varphi-\beta z)],\label{bessel}
\end{eqnarray}
%%%%%%%%%%%%%%%%%%%%%%%%%%%%%%%
where $J_{m}$ is the $m-$th order Bessel function of the first
kind. If the plane waves are coherently superposed without
azimuthal phase modulation ($m=0$), we obtain the bright
beam-like field with the propagation-invariant intensity profile
given by
%%%%%%%%%%%%%%%%%%%%%%%%%%%%
\begin{eqnarray}
I(r,z)=|A_{0}|^{2}J_{0}^{2}(\alpha r).
\end{eqnarray}
%%%%%%%%%%%%%%%%%%%%%%%%%%%%%%%
The radius of the central intensity spot $r_{0}$ is given by
the first zero-point of the Bessel function $J_{0}$ and can be
written as $r_{0}=2.4/\alpha$. In geometrical interpretation that
relation means that the increase of the vertex angle of the conical
surface formed by the propagation vectors of the interfering plane waves
results in the reduction of the size of the intensity spot
of the produced beam. The transverse intensity profile
of the zero-order Bessel beam is illustrated in Fig. 2a.
%%%%%%%%%%%%%%%%%%%%%%%%%%%%%%%
\subsubsection{Nondiffracting vortex beams}
~

During last decade an increasing attention has been focused to
the wavefields possessing the line, spiral or combined wavefront
dislocations. Some types of nondiffracting beams belong to the
class of fields with the spiral wavefront dislocations.
In optics such fields are known as optical vortices.
If the slowly varying complex amplitude $u$ of the nondiffracting
beam (\ref{def}) is rewritten by means of the real amplitude
$u_{0}$ and the phase $\Phi$ in the form
%%%%%%%%%%%%%%%%%%%%%%%%%%%%%%%%
\begin{eqnarray}
u(x,y)=u_{0}(x,y)\exp[i\Phi(x,y)],
\end{eqnarray}
%%%%%%%%%%%%%%%%%%%%%%%%%%%%%%%%
then the point where the phase dislocation appears can be
identified by the nonzero value of the integral
%%%%%%%%%%%%%%%%%%%%%%%%%%%%%
\begin{eqnarray}
\oint_{L}{\nabla\Phi\cdot d{\bf L}},\label{defvort}
\end{eqnarray}
%%%%%%%%%%%%%%%%%%%%%%%%%%%%%
where the integration is performed along the closed line
surrounding the examined point.  The integral (\ref{defvort})
can result in the values $2\pi m$, $m=1,2,\cdots$, where
$m$ represents the topological charge of the vortex field.
At the singularity point the validity of the relations
for real and imaginary parts
$\Re(u)=0$ and $\Im(u)=0$ can be verified
so that we speak about the dark optical vortices.
The simplest type of the nondiffracting vortex beam is the
Bessel beam (\ref{bessel})
represented by the first- or higher-order Bessel
function of the first kind.
The intensity profile of the first-order Bessel beam is
illustrated in Fig. 2b. The helical wavefronts of the Bessel
vortex beams are illustrated in Fig. 3a and 3b for the topological
charges $m=1$ and $m=2$, respectively.
The wavefront dislocations of the optical vortices can be also
identified experimentally applying the interferometric methods.
Interference of the optical vortex possessing the helical
wavefront with the spherical wave results in the typical
spiral patterns. The numerical simulation for the optical vortex
with the topological charges $m=1$ and $m=2$ is
illustrated in  Fig. 4a and 4c.
The pattern obtained by interference of the optical vortex
with the plane wave has the fork-like form. It is illustrated in
Fig. 4b and 4d for the topological charges of the vortex
beam $m=1$ and $m=2$, respectively.
%%%%%%%%%%%%%%%%%%%%%%%%%%%%%%%%
\subsubsection{Mathieu beams}
~

The Bessel beams are obtained if the angular spectrum
(\ref{cas}) consists of the plane waves whose
phases are conveniently modulated and the real amplitudes
remain unmodulated. On the contrary, a good approximation to the
zero-order Mathieu beams can be obtained if the relative phases
of the plane wave components are constant and their
real amplitudes are modified by \cite{vega}
%%%%%%%%%%%%%%%%%%%%%%%%%%%%%%%%
\begin{eqnarray}
A(\psi)=
\exp\left[-\left(\frac{\nu_{0}\cos{\psi}}{w_{0}}\right)^{2}\right],
\end{eqnarray}
%%%%%%%%%%%%%%%%%%%%%%%%%%%%%%%%
where $\nu_{0}$ is given by (\ref{radfr}) and $w_{0}$
denotes the bandwidth of the Gaussian profile
at the plane of spatial frequencies.
The change of the parameter $w_{0}$ results in the change
of the form of the transverse intensity profile of the
nondiffracting Mathieu beam. In Fig. 5a and 5b
the intensity
spots are illustrated for $w_{0}=4\nu_{0}$
and $w_{0}=2\nu_{0}$, respectively.
%%%%%%%%%%%%%%%%%%%%%%%%%%%%%%%%%%%%%%%%%%%
\subsubsection{Caleidoscopic nondiffracting patterns}
~

The function $A$ used in (\ref{cas}) describes the amplitude
and phase modulation of the angular spectrum of the nondiffracting
beam. Usually we assume that it is a continuous function
of the azimuthal angle $\psi$ so that the nondiffracting beam
is produced as the coherent superposition of the plane
waves whose propagation vectors continuously cover the conical surface.
In a special case the nondiffracting field can be obtained
as a discrete superposition of $N$ plane waves.
In that case the modification of the angular spectrum
(\ref{cas}) can be expressed applying the Dirac delta-function
%%%%%%%%%%%%%%%%%%%%%%%%%%%%%%%%
\begin{eqnarray}
A(\psi)=\sum_{j=1}^{N}
A_{0}(\psi)\delta(\psi-\psi_{j}),
\end{eqnarray}
%%%%%%%%%%%%%%%%%%%%%%%%%%%%%%%
where $N$ is an integer. In that case the integration
used in (\ref{integ}) is replaced by the summation and
the complex amplitude of the
nondiffracting field can be rewritten as
%%%%%%%%%%%%%%%%%%%%%%%%%%%%%%%%%%
\begin{eqnarray}
a(r,\varphi,z)=\frac{ik}{2\pi}
\exp(-i\beta z)\sum_{j=1}^{N}
A_{0}(\psi_{j})
\exp[i\alpha r\cos(\psi_{j}-\varphi)].\label{sum}
\end{eqnarray}
%%%%%%%%%%%%%%%%%%%%%%%%%%%%%%%
The numerical simulation of the transverse intensity
profile of the nondiffracting beam is illustrated in Fig. 6.
The azimuthal angles $\psi_{j}$ related to the interfering plane
waves are chosen as
%%%%%%%%%%%%%%%%%%%%%%%%%%%%%%%
\begin{eqnarray}
\psi_{j}=(j-1)\Delta\psi,
\end{eqnarray}
%%%%%%%%%%%%%%%%%%%%%%%%%%%%%%%
where $\Delta\psi=2\pi/N$ and $A_{0}$ is assumed
to be a constant. The intensity spots illustrated in
Fig. 6a-d are obtained for $N=5,\  10,\  15$ and $30$,
respectively. If the number of interfering plane waves is
even $N=2N'$ and the angular spectrum is modified by the
function possessing property $A_{0}(\psi)=A_{0}(\psi+\pi)$
the complex amplitude of the nondiffracting field (\ref{sum})
can be alternatively expressed as a superposition of the
azimuthally rotating cosine gratings
%%%%%%%%%%%%%%%%%%%%%%%%%%%%%%%
\begin{eqnarray}
a(r,\varphi,z)=\frac{ik}{\pi}
\exp(-i\beta z)\sum_{j=1}^{N'}
A_{0}(\psi_{j})
\cos[\alpha x\cos(\psi_{j})+\alpha y \sin(\psi_{j})].
\end{eqnarray}
%%%%%%%%%%%%%%%%%%%%%%%%%%%%%%%
The nondiffracting pattern Fig.6b obtained due to the
interference of $N=10$ plane waves possessing the same amplitude
and phase can also be created as a coherent superposition
of $N'=5$ cosine gratings rotating in the azimuthal
direction with the angular increment $\Delta\psi=\pi/N'$.
The cosine grating components  and the transverse intensity
pattern obtained by their superposition are illustrated in
Fig. 7. The nondiffracting field described by the integral
(\ref{integ}) can also be interpreted as the superposition
of the cosine gratings. In that case the gratings
are continuously rotating along the azimuthal angle
\cite{bouchal3}.
%%%%%%%%%%%%%%%%%%%%%%%%%%%%%%%
\section{PARTIALLY-COHERENT NONDIFFRACTING BEAMS}
~
In the case of the fully coherent nondiffracting beam the
azimuthal amplitude and phase modulation of the angular spectrum
is described by the deterministic function $A$. As has been shown
its changes can be used for the formation of the transverse
intensity profile of the beam.  The partially-coherent nondiffracting
beam is still related to the angular spectrum
(\ref{cas}) but the modulation function $A$
is given as a product of the deterministic amplitude $A_{d}$
and the random amplitude $A_{r}$,
%%%%%%%%%%%%%%%%%%%%%%%%%%%%%%%%%
\begin{eqnarray}
A(\psi)=A_{d}(\psi)A_{r}(\psi).\label{ranamp}
\end{eqnarray}
%%%%%%%%%%%%%%%%%%%%%%%%%%%%%%%
In that case the partially-coherent nondiffracting beam is
described by the second-order cross-spectral density function
%%%%%%%%%%%%%%%%%%%%%%%%%%%%%%%%%
\begin{eqnarray}
W({\bf r_{1}},{\bf r_{2}})=<a^{*}({\bf r_{1}})a({\bf r_{2}})>,
\end{eqnarray}
%%%%%%%%%%%%%%%%%%%%%%%%%%%%%%%%%
where ${\bf r_{j}}\equiv (x_{j},y_{j},z_{j})$ is the position
vector. Applying (\ref{integ}), (\ref{cas}) and (\ref{ranamp})
we obtain
%%%%%%%%%%%%%%%%%%%%%%%%%%%%%%%%
\begin{eqnarray}
W({\bf r_{1}},{\bf r_{2}})&=&\left(\frac{k}{2\pi}\right)^{2}
\int_{-\pi}^{\pi}\int_{-\pi}^{\pi}
\gamma(\psi_{1},\psi_{2})|A_{r}(\psi_{1})||A_{r}(\psi_{2})|
A^{*}_{d}(\psi_{1})A_{d}(\psi_{2})\cr
&\times&f^{*}({\bf r_{1}},\psi_{1})f({\bf r_{2}},\psi_{2})
d\psi_{1}d\psi_{2},\label{csdf}
\end{eqnarray}
%%%%%%%%%%%%%%%%%%%%%%%%%%%%%%%%%
where $\gamma$ is the degree of angular correlation defined as
%%%%%%%%%%%%%%%%%%%%%%%%%%%%%%
\begin{eqnarray}
\gamma(\psi_{1},\psi_{2})=
\frac{<A^{*}_{r}(\psi_{1})A_{r}(\psi_{2})>}
{|A_{r}(\psi_{1})||A_{r}(\psi_{2})|},
\end{eqnarray}
%%%%%%%%%%%%%%%%%%%%%%%%%%%%%%%%%
and $< >$ means ensemble average over field realizations.
It describes the mutual correlation of the plane wave components
of the angular spectrum propagating with the propagation vectors
${\bf k_{1}}$ and ${\bf k_{2}}$. The propagation vectors create the
conical surface with the vertex angle $\theta_{0}$ so that
their longitudinal $z-$components are constant.
The transverse
$x-$ and $y-$components depend on the azimuthal angle $\psi$.
The propagation vector ${\bf k_{j}}$ then possesses components
%%%%%%%%%%%%%%%%%%%%%%%%%%%%
$(|{\bf
k_{j}}|\sin{\theta_{0}}\cos{\psi_{j}},
|{\bf
k_{j}}|\sin{\theta_{0}}\sin{\psi_{j}},
|{\bf k_{j}}|\cos{\theta_{0}})$.
%%%%%%%%%%%%%%%%%%%%%%%%%%%%%%%%%%
The simplest type of the partially-coherent nondiffracting
field is obtained if the plane wave components of the angular
spectrum are superposed incoherently so that $\gamma$
is given by the Dirac delta function
%%%%%%%%%%%%%%%%%%%%%%%%%
\begin{eqnarray}
\gamma(\psi_{1},\psi_{2})=\delta(\psi_{1}-\psi_{2}).
\end{eqnarray}
%%%%%%%%%%%%%%%%%%%%%%%%%
If the deterministic function $A_{d}$ modulates
only the phase of the angular spectrum the cross-spectral
density function (\ref{csdf}) becomes
%%%%%%%%%%%%%%%%%%%%%%%%%%%%%%%%%
\begin{eqnarray}
W({\bf r_{1}},{\bf r_{2}})=C\exp[-i\beta(z_{2}-z_{1})]
J_{0}(\alpha \rho),\label{csde}
\end{eqnarray}
%%%%%%%%%%%%%%%%%%%%%%%%%%%%%%%%%%
where
%%%%%%%%%%%%%%%%%%%%%%%%%%%%%%%%%%
\begin{eqnarray*}
\rho=[(x_{1}-x_{2})^{2}+(y_{1}-y_{2})^{2}]^{1/2},
\end{eqnarray*}
%%%%%%%%%%%%%%%%%%%%%%%%%%%%%%%%%%%
and $C$ denotes the constant amplitude.
%%%%%%%%%%%%%%%%%%%%%%%%%%%%%%%%%%%
The cross-spectral density function (\ref{csde}) describes the
Bessel correlated field with the constant intensity and the
sharply peaked transverse-spatial-correlation profile
\cite{turunen}. The more general partially-coherent
nondiffracting fields are examined in \cite{bouchal6}.
They are obtained on the assumption that the degree
of angular correlation $\gamma$ can be controlled.
The controlled change of the mutual correlation
of the plane wave components of the angular spectrum
offers a possibility to change the transverse intensity
profile of the beam to the desired form and to control
the coherence properties of the generated nondiffracting
beam. The controlled change of $\gamma$ can be
realized experimentally in the illumination chain using
the pseudo-thermal spatially incoherent source
such as the Gaussian Shell-model source \cite{mandel}.
Analysis of that
experimental set-up is presented in \cite{bouchal6}.
The change of the transverse
intensity profile of the partially-coherent nondiffracting
beam caused by the change of the degree of angular correlation
is illustrated in Fig. 8. The intensity spot in Fig. 8a is obtained
for the coherent superposition of the plane wave components
$(\gamma=1)$. The intensity patterns in Figs. 8b, 8c and 8d
are obtained by the partially-coherent superposition
of the plane waves. They are related to the
functions $\gamma$ illustrated in
Fig. 9 by the curves denoted as '$\bullet\bullet\bullet$',
'ooo', and '$-$',
respectively. The coherent beam with the intensity profile
in Fig. 8a represents the nondiffracting vortex of the Bessel
type possessing the topological charge $m=2$.
The beam is dark (axial intensity is equal to zero)
and its wavefront has the helical form. In the interference
experiment using the spherical reference wave
the phase singularity is visualized by the spiral
interference pattern in Fig. 10a. Due to the change of the spatial
coherence the beam becomes bright (axial intensity is nonzero)
and the phase dislocation is removed. Thit is obvious from
Figs. 10b-d illustrating the vanishing spiral character
of the interference pattern.
%%%%%%%%%%%%%%%%%%%%%%%%%%%%%%%%%%%
\section{SPATIAL SHAPING OF NONDIFFRACTING FIELDS}
~
The condition of nondiffracting beam propagation (\ref{cas})
puts restrictions only on the propagation directions of the plane wave
components of the angular spectrum. Their amplitudes and phases
are given by $A(\psi)$ and can be arbitrary. The freedom of the
amplitude and the phase modulation of the angular spectrum can be
applied to the control of the spatial shaping of the produced
nondiffracting patterns. In \cite{bouchol,boukyvn}, the azimuthal
modulation of the form $A(\psi)=t(\psi)a(\psi)$ is assumed.
The function $t$ is used to control the shape of the created
nondiffracting spot and $a$ is applied to move it to the required
position at the transverse plane. Furthermore, the size of the
nondiffracting spot can be changed by the parameter $\alpha$
defined by (\ref{radfr}). The shift of the centre of the
nondiffracting spot to the point with coordinates $\Delta x$
and $\Delta y$ is achieved if the function $a$ is of the form
%%%%%%%%%%%%%%%%%%%
\begin{eqnarray}
a(\psi)=W(\Delta x,\Delta y)
\exp[i\alpha(\Delta x\cos{\psi}+\Delta y\sin{\psi})],
\end{eqnarray}
%%%%%%%%%%%%%%%%%%%
where $W$ is the weighing function representing the amplitude of
the nondiffracting spot at the position $(\Delta x,\Delta y)$.
The nondiffracting pattern whose transverse amplitude profile is
composed of the continuously shifted nondiffracting spots is then
described by the convolution
%%%%%%%%%%%%%%%%%%%
\begin{eqnarray}
U({\bf r})=\exp[i(\omega t-\beta z)]
\int\int_{-\infty}^{+\infty}W(\Delta x,\Delta y) T(x-\Delta
x,y-\Delta y)d\Delta x d\Delta y,
\end{eqnarray}
%%%%%%%%%%%%%%%%%%%
where
%%%%%%%%%%%%%%%%%%%
\begin{eqnarray}
T(x-\Delta x,y-\Delta y)=\int_{0}^{2\pi}
t(\psi)\exp\{-i\alpha[(x-\Delta x)\cos{\psi}
+(y-\Delta y)\sin{\psi}\}d\psi.
\end{eqnarray}
%%%%%%%%%%%%%%%%%%%
By the control of amplitudes, shapes, sizes and shifts of the
nondiffracting spots we can create the total field whose
transverse amplitude profile approximates the form predetermined
by the function $W$.
If the nondiffracting spots are superposed coherently the
similarity between the predetermined and the realized amplitude
profiles is degraded by the interference effects. It can be
significantly improved if the nondiffracting spots are mutually
incoherent.
In \cite{bouchol}, the simple
experimental implementation of the controllable spatial shaping of
the nondiffracting fields was proposed and realized for both the
coherent and incoherent light. The obtained results are shown
in Figs. 11 and 12. In Fig. 11a, the predicted transverse profile
defined by $W$ is illustrated. It has the form of the array
composed of the separated point sources. The transverse
intensity profiles of the generated nondiffracting beams obtained
with incoherent and coherent light are shown in Figs. 11b and 11c,
respectively.
In both cases the generated fields are
nondiffracting so that the intensity profiles do not change under
propagation.
The intensity profiles illustrated in Figs. 11d and 11e are again
obtained for incoherent and coherent light but the parameter
$\alpha$ is change in such a way that the size of the
nondiffracting spots is reduced so that their resolution is
improved.
As is obvious, the similarity between the required
and generated intensity profiles is much more better for
incoherent light than for coherent one. In Fig. 12 the similar
situation is illustrated for the required profile resembling the initials of
Palack\'y University. In that case, the transverse intensity
profile is obtained as a continuous superposition of the
nondiffracting spots.

%%%%%%%%%%%%%%%%%%%%%%%%%%%%%%%%%%%
\section{VECTORIAL NONDIFFRACTING BEAMS}
~
The detailed study of the nondiffracting beams including analysis
of the polarization states and the flow of the electromagnetic energy
requires the vectorial electromagnetic description.
The problem is to find the monochromatic electromagnetic field
whose vector complex amplitudes exactly fulfil the Maxwell
equations and can be written in the form
%%%%%%%%%%%%%%%%%%%%%%%%%%%%%%%%%%%
\begin{eqnarray}
{\bf E}(x,y,z,t)&=&{\bf e}(x,y)\exp[i(\omega t-\beta
z)],\label{vnb1}\\
{\bf H}(x,y,z,t)&=&{\bf h}(x,y)\exp[i(\omega t-\beta
z)],\label{vnb2}
\end{eqnarray}
%%%%%%%%%%%%%%%%%%%%%%%%%%%%%%%%%%%
where ${\bf e}$ and ${\bf h}$ are the propagation invariant
amplitudes.
In \cite{romea} the special type of the vectorial nondiffracting
beams was introduced on the assumption that the longitudinal
component of the electric field resembles the zero-order Bessel
function of the first kind. The model resulted in a radially
polarized beam whose radial electric field corresponds to the
first-order Bessel function of the first kind.
In \cite{bouchal2} the representative theorem for the Helmholtz
equation was applied to derive ${\bf E}$ and ${\bf H}$
from the scalar complex amplitudes $a_{m}$ exactly
fulfilling the scalar Helmholtz equation. The vector complex
amplitudes then can be written as
%%%%%%%%%%%%%%%%%%%%%%%%%%%%%%%%%%
\begin{eqnarray}
{\bf E}&=&
-\sum_{m}(p_{m} {\bf P}_{m}+q_{m}{\bf Q}_{m})
\exp(i\omega t),\label{vec1}\\
{\bf H}&=&i\xi
\sum_{m}(p_{m} {\bf Q}_{m}+q_{m}{\bf P}_{m})
\exp(i\omega t),\label{vec2}
\end{eqnarray}
%%%%%%%%%%%%%%%%%%%%%%%%%%%%%%%%%%%
where
%%%%%%%%%%%%%%%%%%%%%%%%%%%%%%%%%%%
\begin{eqnarray}
{\bf P}_{m}&=&-({\bf s}\times\nabla a_{m}),\cr
{\bf Q}_{m}&=&\frac{1}{k}\nabla\times {\bf P}_{m},
\end{eqnarray}
%%%%%%%%%%%%%%%%%%%%%%%%%%%%%%%%%%%
and $1/\xi=\sqrt{\mu_{0}/\epsilon_{0}}$ is the impedance of the
vacuo.
%%%%%%%%%%%%%%%%%%%%%%%%%%%%%%%%%%%
It can be shown that the vector amplitudes (\ref{vec1})
and (\ref{vec2}) exactly fulfil the Maxwell equations and can be
expressed in the form (\ref{vnb1}) and (\ref{vnb2}).
They are constructed from the base of the scalar nondiffracting
beams.
The complex amplitudes $a_{m}$ represent the independent
scalar nondiffracting solutions to the Helmholtz equation
obtained for the same angular wavenumber $\beta$.
In the vectorial solution they are applied with the weighing
coefficients $p_{m}$ and $q_{m}$.
The used symbols $\epsilon$, $\mu$ and ${\bf s}$
denote the permittivity, the permeability and an arbitrary
constant vector, respectively. The nondiffracting fields related
to the single summation indices represent the
nondiffracting modes. The vector nondiffracting fields
can then be constructed as the one- or multi-mode fields.
Classification and analysis of their properties are presented
in \cite{horak}. The special types of the vector nondiffracting
beams are obtained by an appropriate choice of the weighing
coefficients $p_{m}$ and $q_{m}$ and the constant vector ${\bf s}$.
For example, the transversal electric (TE) field follows from
(\ref{vec1}) and (\ref{vec2}) used
with $p_{m}\not =0$, $q_{m}=0$ and ${\bf s}\equiv (0,0,s_{z})$.
The transversal magnetic (TM) field is obtained with
$p_{m}=0$, $q_{m}\not =0$ and ${\bf s}\equiv (0,0,s_{z})$.
If the complex amplitudes of the scalar nondiffracting beams
$a_{m}$ are represented by the Bessel functions the vector
nondiffracting beams possessing the azimuthal or the radial
polarization can be obtained. In that case,
the vector complex amplitudes of the electric and the magnetic fields
can be conveniently expressed by the radial, azimuthal and
longitudinal components ${\bf E}\equiv (E_{r},E_{\varphi},E_{z})$
and ${\bf H}\equiv (H_{r},H_{\varphi},H_{z})$. The nondiffracting
TE beam
possessing the azimuthal polarization of the electric field
can be described by the vectors ${\bf E}\equiv (0,E_{\varphi},0)$
and ${\bf H}\equiv (H_{r},0,H_{z})$. Their components can be
written as \cite{bouchal2}
%%%%%%%%%%%%%%%%%%%%%%%%%%%%%%%%%%%%
\begin{eqnarray}
E_{r}&=&0,\\
E_{\varphi}&=&-p_{0}\alpha J_{1}(\alpha r)\exp[i(\omega t-\beta z)],\\
E_{z}&=&0,\\
H_{r}&=&-p_{0}\alpha\xi\frac{\beta}{k}J_{1}(\alpha r)
\exp[i(\omega t-\beta z)],\\
H_{\varphi}&=&0,\\
H_{z}&=&ip_{0}\xi\frac{\alpha^{2}}{k}J_{0}
(\alpha r)\exp[i(\omega t-\beta z)].
\end{eqnarray}
%%%%%%%%%%%%%%%%%%%%%%%%%%%%%%%%%%%%
In the case of the TM beam the magnetic field possesses the
azimuthal polarization while the electric field is radially
polarized. The components of the field vectors can be expressed
as
%%%%%%%%%%%%%%%%%%%%%%%%%%%%%%%%%%
\begin{eqnarray}
E_{r}&=&-iq_{0}\alpha\frac{\beta}{k}
J_{1}(\alpha r)\exp[i(\omega t-\beta z)],\\
E_{\varphi}&=&0,\\
E_{z}&=&-q_{0}\frac{\alpha^{2}}{k}J_{0}(\alpha r)
\exp[i(\omega t -\beta z)],\\
H_{r}&=&0,\\
H_{\varphi}&=&iq_{0}\xi\alpha J_{1}(\alpha r)
\exp[i(\omega t -\beta z)],\\
H_{z}&=&0.
\end{eqnarray}
%%%%%%%%%%%%%%%%%%%%%%%%%%%%%%%%%%%%
The azimuthally and the radially polarized fields can be considered
as the linearly polarized fields with the spatial change
of the direction of oscillations. At each point of the
transverse plane the field vectors
are linearly polarized. For the radially polarized field
the direction of oscillations
is given by the line connecting that point with the center of the
beam while the azimuthally polarized field oscillates along the
direction orthogonal to that line. The TE nondiffracting beam
with the azimuthal polarization of the electric field and the
radially polarized magnetic field is illustrated in Fig. 13.
In Fig. 13a the magnitude and the direction
of the transverse component of the electric intensity
${\bf E}_{\perp}\equiv (E_{r},E_{\varphi})$ are illustrated by
the arrows at the separate points of the transverse plane.
The radially polarized magnetic field is similarly illustrated
in Fig. 13b.
%%%%%%%%%%%%%%%%%%%%%%%%%%%%%%%%%%%%
\section{PROPERTIES OF NONDIFFRACTING BEAMS}
~

The nondiffracting beams exhibit interesting properties by which
they differ from the common types of beams, for example Gaussian
beams. Recently, an increasing attention was devoted to those
properties because
they offer many potential applications and their explanation
can be important for better understanding of the origin of the
diffraction phenomena and of the nature of the electromagnetic
field. Some of them are briefly discussed.
%%%%%%%%%%%%%%%%%%%%%%%%%%%%%%%%%%%%%
%subsection{Healing effect}
\subsection{Beam robustness}
~
An important property of the nondiffracting beam is its
resistance against amplitude and phase distortions.
The transverse intensity profile of the nondiffracting beam
disturbed by the nontransparent obstacle regenerates during
free-propagation behind that obstacle.
The healing effect causes that the initial transverse intensity
profile is restored certain distance behind the obstacle.
The effect was explained theoretically applying Babinet's
principle and verified experimentally \cite{chlup}.
The results of the experiment are presented in Fig. 14.
The nondiffracting beam with
the transverse intensity profile approximately corresponding to the
first-order Bessel function $J_{1}$ is disturbed by the
nontransparent rectangular obstacle. During free-propagation
of the beam behind the obstacle its transverse intensity profile
regenerates. As is obvious from Fig. 14, the initial
Bessel-like profile is restored with a very good fidelity.
%%%%%%%%%%%%%%%%%%%%%%%%%%%%%%%%%%%%%
\subsection{Beam energetics}
~

In the framework of the Maxwell theory the energetics of the
optical beams is characterized by the density of the flow of the
electromagnetic energy denoted as the Poynting vector ${\bf S}$ and by the
volume density of the electromagnetic energy $w$. Both quantities
are spatially and temporally dependent and are mutually coupled
in the energy conservation law. In nonconducting media it can be
expressed by
%%%%%%%%%%%%%%%%%%%%%%%%%%%%%%%%%%%%
\begin{eqnarray}
\nabla\cdot {\bf S}+\frac{\partial w}{\partial
t}=0.\label{zakzach}
\end{eqnarray}
%%%%%%%%%%%%%%%%%%%%%%%%%%%%%%%%%%%%%
In real situations the temporally averaged quantities
$<{\bf S}>$ and $<w>$ are of the particular importance.
For the monochromatic nondiffracting electromagnetic beam
(\ref{vnb1}) and (\ref{vnb2}) they can be written as
%%%%%%%%%%%%%%%%%%%%%%%%%%%%%%%%%%%%%
\begin{eqnarray}
<{\bf S}>&=&{\bf e}^{*}\times{\bf h}+{\bf e}\times{\bf
h}^{*},\label{Poynting}\\
<w>&=&\epsilon({\bf e}^{*}\cdot {\bf e})+\mu({\bf h}^{*}\cdot {\bf
h}).
\end{eqnarray}
%%%%%%%%%%%%%%%%%%%%%%%%%%%%%%%%%%%%%
As the vector amplitudes ${\bf e}$ and ${\bf h}$ of the
nondiffracting beams are independent
of the $z-$coordinate, the energy conservation law can be
simplified to the form
%%%%%%%%%%%%%%%%%%%%%%%%%%%%%%%%%%%%%%
\begin{eqnarray}
\nabla\cdot <{\bf S}_{\perp}>=0,\label{eclnb}
\end{eqnarray}
%%%%%%%%%%%%%%%%%%%%%%%%%%%%%%%%%%%%%%
where $<{\bf S}_{\perp}>$ denotes the transverse part of the
Poynting vector. The transverse and the longitudinal components
of the Poynting vector fulfil relations $<{\bf S}_{\perp}>\cdot\
{\bf z}=0$ and ${\bf S}_{\parallel}\times {\bf z}=0$, where
${\bf z}$ is the unit vector coinciding with the $z$-axis.
The Poynting vector providing the density of the electromagnetic
energy flow can be obtained as the vector sum
$<{\bf S}>=<{\bf S}_{\perp}>+<{\bf S}_{\parallel}>$.
The energy conservation law formulated for the nondiffracting
beams (\ref{eclnb}) requires the zero divergence of the
transverse component of the Poynting vector but the transverse
energy flow itself can be nonzero. The fact that the
monochromatic nondiffracting beam whose transverse intensity profile
remains unchanged under propagation can exhibit the nonzero
energy flow orthogonal to the direction of propagation is
surprising. It excited interest in the structure of the
transverse component of the Poynting vector.
Applying the system of the circular cylindrical coordinates
it can be decomposed into the radial and azimuthal components
$<{\bf S}_{\perp}>=<{\bf S}_{r}>+<{\bf S}_{\varphi}>$. It can be
shown that for the one-mode nondiffracting beam the radial flow must
be equal to zero and the transverse energy flow can possess only
azimuthal component. Due to the superposition with the longitudinal
component of the Poynting vector the total flow has the helical
character. In a general case of the multi-mode nondiffracting
beam the radial energy flow can be nonzero due to the
interference of the modes. As is obvious from (\ref{eclnb})
the points of the transverse plane where the transverse energy
flow has the constant magnitude lye on the closed lines.
An exhaustive analysis of the energetic properties of the vector
nondiffracting beams is presented in \cite{horak}.
%%%%%%%%%%%%%%%%%%%%%%%%%%%%%%%%%%%%%%%%
%\subsection{Transfer of angular momentum}
\subsection{Orbital angular momentum}
~

The nondiffracting vortex beams carry the angular momentum
which can be transferred to atoms and microscopical particles.
The mechanical consequence of that interaction is rotation of the
particles.
The angular momentum has two components - the orbital angular
momentum and the spin.  The spin depends on the polarization
state of the beam and is equal to zero for the linear
polarization. If the beam interacts with particles, the spin
causes rotation of the particles around their own axis.
The orbital angular momentum is a consequence of the
spiral flow of the electromagnetic energy and is typical for
beams with the helical wavefront. They are known as the vortex beams.
Under interaction with the particle, the orbital angular momentum
causes its rotation around the center of the vortex nested in the host
beam.

The angular momentum ${\bf J}$ can be written as a vectorial product of the
position vector ${\bf r}$ and the linear momentum ${\bf p}$,
%%%%%%%%%%%
\begin{eqnarray}
{\bf J}={\bf r}\times{\bf p}.
\end{eqnarray}
%%%%%%%%%%%%
If the beam propagates in vacuo with velocity $c$,
${\bf p}$ can be expressed by means of the Poynting vector as
%%%%%%%%%%%
\begin{eqnarray}
{\bf p}=\frac{{\bf S}}{c^{2}}.
\end{eqnarray}
%%%%%%%%%%%
Assuming the beam propagating along the $z$-direction, its
orbital angular momentum is given by the $z$-component of ${\bf J}$,
%%%%%%%%%%%
\begin{eqnarray}
J_{z}=\frac{({\bf r}\times{\bf S})_{z}}{c^{2}}.
\end{eqnarray}
%%%%%%%%%%%
Applying the circular cylindrical coordinates $r,\varphi,z$, we obtain
%%%%%%%%%%
\begin{eqnarray}
J_{z}=\frac{rS_{\varphi}}{c^2},\label{defjz}
\end{eqnarray}
%%%%%%%%%%%
where $S_{\varphi}$ is the magnitude of the azimuthal component
of the Poynting vector.
Performing normalization of the orbital angular momentum by
the volume energy density,
%%%%%%%%%%
\begin{eqnarray}
j_{z}=\frac{J_{z}}{w},
\end{eqnarray}
%%%%%%%%%%
the quantity $j_{z}$ can be interpreted as a magnitude of the
orbital angular momentum carried by the photon of the beam.
%%%%%%%%%%%%%%%%%%%%%%%%%%%%%%%%%%%%%%%%
%\subsubsection{Orbital angular momentum}
%~

The orbital angular momentum of the optical beam can be expressed
by the Poynting vector representing the density of the flow of the
electromagnetic energy. In the exact vectorial theory,
the Poynting vector is given by the vectors of the
electromagnetic field (\ref{Poynting}) and fulfils the energy
conservation law (\ref{zakzach}).
In the framework of the scalar approximation the optical
beams are  described by the scalar complex amplitude
$U$ fulfilling the wave equation. By simple manipulations of the
wave equation, the relation resembling the form of the energy
conservation law can be obtained
%%%%%%%%%%%%%%%%
\begin{eqnarray}
\nabla\cdot {\bf S'}+\frac{\partial w'}{\partial t}=0,
\end{eqnarray}
%%%%%%%%%%%%%%%%%%%%%%%%%%%%%%%%%%%%%%%%
where
%%%%%%%%%%%%%%%%%%%%%%%%%%%%%%%%%%%%%%%%
\begin{eqnarray}
{\bf S'}&=&-\left(\nabla U\frac{\partial U^{*}}
{\partial t}+\nabla U^{*}\frac{\partial U}
{\partial t}\right),\label{appoynt}\\
w'&=&\nabla U\cdot\nabla U^{*}+\frac{1}{c^{2}}
\left|\frac{\partial U}{\partial t}\right|^{2}.
\end{eqnarray}
%%%%%%%%%%%%%%%%%%%%%%%%%%%%%%%%%%%%%%%%
The quantities ${\bf S'}$ and $w'$ can be comprehended as
approximations to the Poynting vector and the volume energy
density of the optical beam, respectively.
The orbital angular momentum can then be simply demonstrated on
the case of the scalar monochromatic vortex beam. If the vortex
is nested at the centrum of the nondiffracting host beam, its complex
amplitude can be expressed by means of the circular cylindrical
coordinates as
%%%%%%%%%%%%%%%%%%%%%%%%%%%%%%%%
\begin{eqnarray}
U(r,\varphi,z,t)=a(r)\exp[i(\omega t+m\varphi-\beta z)],
\end{eqnarray}
%%%%%%%%%%%%%%%%%%%%%%%%%%%%%%%%%%%%%%%%
where $m$ denotes the topological charge of the vortex and $\beta$
is the angular wavenumber of the host nondiffracting beam.
%%%%%%%%%%%%%%%%%%%%%%%%%%%%%%%%%%%%%%%%
If $S_{\varphi}$ used in (\ref{defjz}) is  replaced by its
scalar approximation $S'_{\varphi}$ following from
(\ref{appoynt}), we obtain
%%%%%%%%%%%%%%%%%%%%%
\begin{eqnarray}
J_{z}=\frac{2m\omega|a|^{2}}{c^{2}}.
\end{eqnarray}
%%%%%%%%%%%%%%%%%%%%%%%%%%%%%%%%%%%%%%%%
Scalar approximation to the volume energy density can be
expressed as
%%%%%%%%%%%%%%
\begin{eqnarray}
w'=2\beta^{2}|a|^{2}.
\end{eqnarray}
%%%%%%%%%%%%%%%%%%%%%%%%
The orbital angular momentum carried by the single photon of the
monochromatic vortex beam with the angular frequency $\omega$ is
then given by
%%%%%%%%%%%%%%%%%%%%%%
\begin{eqnarray}
j_{z}=\frac{m}{\omega}.
\end{eqnarray}
%%%%%%%%%%%%%%%%%%%%%%%%
The orbital angular momentum is proportional to the topological
charge of the vortex beam and inversely proportional to its
angular frequency. It is a reason why the
mechanical influence of the vortex is more appreciable for the
microwaves than for the optical waves.

One of the most important tasks in design of MEMS
(Micro Electro Mechanical Systems) is to find ways how to power
machines that measure only microns across. The promising solution
is to rotate them by the blowing of "the light wind". It can be
produced by the optical vortices carrying the orbital angular
momentum. In \cite{bn1,bn2}, the simple model of the vortex beam
interaction accompanied by the exchange of the orbital angular
momentum was proposed and analyzed. It was shown that both the
phase topology and the local distribution of the orbital angular
momentum of the vortex nested in the nondiffracting host beam can
revive and regenerate to the initial form after interaction with
the 2D object. In the simulation model, the rotationally
nonsymmetrical object takes the orbital angular momentum from the
beam and rotates. The healing of the spatial distribution of the
orbital angular momentum after interaction is illustrated in Fig.
15. It is accompanied by the self-regeneration of the phase
topology of the vortex beam. In Figs. 16a and 16b, the
vortex helical wavefront of the initial beam  is
visualized by interference with  plane and spherical waves,
respectively. After interaction with 2D object, the phase topology
is strongly disturbed  but during free
propagation revives to the initial form (Figs. 16c-16f).
%%%%%%%%%%%%%%%%%%%%%%%%
%\subsubsection{Spin}
%~
%
%For demonstration of the spin, the polarization properties of the
%beam must be included so that the exact vectorial description is
%required. It can be performed applying the electromagnetic
%potentials. The vectors of the electromagnetic field ${\bf E}$
%and ${\bf B}$ then can be expressed by means of the vector
%potential ${\bf A}$ and the scalar potential $\Phi$ in the form
%%%%%%%%%%%%%%%%%%%%%%%%%%%%%%%%%%%%%%%%
%\begin{eqnarray}
%{\bf B}&=&\nabla\times{\bf A},\\
%{\bf E}&=&-\frac{\partial {\bf A}}{\partial t}-\nabla \Phi.
%\end{eqnarray}
%%%%%%%%%%%%%%%%%%%%%%%%%%%%%%%%%%%%%%%%
\subsection{Transversality of electric and magnetic field}
~

An ideal homogeneous monochromatic plane wave
is the transversal electromagnetic (TEM) wave.
Its electric and magnetic fields
oscillate at the transverse planes so that the projections
of the vectors ${\bf E}$ and ${\bf H}$ to the direction of
propagation are equal to zero.
For the free-space propagation that property follows
directly from the Maxwell equations $\nabla\cdot{\bf E}=0$
and $\nabla\cdot{\bf H}=0$.
In real cases the wave is not
homogeneous (vectors ${\bf E}$ and ${\bf H}$ depend on the
transverse coordinates) so that the electromagnetic transversality
cannot be exactly achieved. Nevertheless, the common beams,
for example Gaussian beams, can be considered to be nearly
transversal with respect to the dominant propagation direction.
That means that the longitudinal component of the electric or
magnetic vector of the beam is very small in comparison with
the transversal one. Applying the concept of the nondiffracting
propagation we can prepare quite different situation when the
longitudinal component of the electric or magnetic filed is
comparable to the transverse component. That property is given
by the structure of the spatial spectrum of the nondiffracting
beam and depends on the beam spot size. It was verified that the
extremely strong longitudinal component of the electric field of
the nondiffracting beam can be obtained only if the transverse
dimensions of the beam are comparable to the wavelength
\cite{bouchal2}. The beams with the strong longitudinal field
are important for applications and were successfully applied to
the design of the electron accelerators \cite{romea}.
%%%%%%%%%%%%%%%%%%%%%%%%%%%%%%%%%%%%%%
\subsection{Self-imaging}
~

The basic parameter of the scalar nondiffracting beam (\ref{def})
is the angular wavenumber $\beta$. If we perform the coherent
superposition of two nondiffracting modes possessing the different
angular wavenumbers $\beta_{1}$ and $\beta_{2}$,
the produced beam is not propagation invariant. Its transverse
intensity profile depends on the $z$-coordinate and reappears
periodically in the free-space propagation. In dependence on the
propagation coordinate the beam axial intensity changes
sinusoidally.
%The longitudinal period of the change can be
%expressed as $L=2\pi n/(\beta_{1}-\beta_{2})$, $n=1,2,\cdots$.
Under convenient choice of the angular wavenumbers that effect
can be realized also by the coherent superposition of $m$ nondiffracting
beams.
The effect represents the spatial analog of the mode-locking realized in
the temporal domain. Due to the interference of the modes, the
transverse intensity profile of the beam reappears periodically
at the planes of the constructive interference and vanishes at
the places of the destructive interference. Its complex amplitude
$a$ then fulfils condition
%%%%%%%%%%%%%%%%%%%%%%%%%%%%%%%%
\begin{eqnarray}
a(x,y,z)=a(x,y,z+L),
\end{eqnarray}
%%%%%%%%%%%%%%%%%%%%%%%%%%%%%%%%
where $L$ is the longitudinal period.
%%%%%%%%%%%%%%%%%%%%%%%%%%%%%%%%
The beam axial
intensity is significantly nonzero only near the planes of the
constructive interference and vanishes elsewhere.
The intensity distribution of the field exhibiting the
self-imaging effect is illustrated in Fig. 17.
It can be obtained as the coherent superposition of the
nondiffracting modes whose angular wavenumbers are adapted to
the chosen period $L$ as
%%%%%%%%%%%%%%%%%%%%%%%%%%
\begin{eqnarray}
\beta_{m}=2m\pi/L,\ \ \ m=0,1,2,\cdots,\ \ m<L/\lambda.
\label{sic}
\end{eqnarray}
%%%%%%%%%%%%%%%%%%%%%%%%%%
The width of the peaks of the axial intensity $I(0,0,z)$ can be
decreased if the number of the superposed modes is increased.
The width of the transverse intensity profile decreases with the
increasing values of the angular wavenumbers of the
used nondiffracting modes.
The general vectorial treatment of the self-imaging effect and
its experimental verification was presented in \cite{bouchal4}.
%%%%%%%%%%%%%%%%%%%%%
\subsection{Self-reconstruction ability}
~

The beams possessing the self-reconstruction ability belong
to the class of fields exhibiting the longitudinal propagation
periodicity. The Talbot effect and the self-imaging are also
members of that class of fields. To express their distinctions
the exact definition of the self-reconstruction effect is
necessary. It can be introduced applying the concept of the
shape-invariant transformation. The spatial evolution of the
complex amplitude of the monochromatic field fulfilling the
Helmholtz equation can be expressed by the integral operator
$\Gamma$ as
%%%%%%%%%%%%%%%%%%%%%%%%%%%%%%%
\begin{eqnarray}
a(x,y,z)=\Gamma a_{0}(x,y,z_{0}),
\end{eqnarray}
%%%%%%%%%%%%%%%%%%%%%%%%%%%%%%
where $a_{0}$ is the complex amplitude at the $z=z_{0}$ plane.
The shape-invariant transformation is achieved if the complex
amplitude can be written in the form
%%%%%%%%%%%%%%%%%%%%%%%%%%%%%%%%
\begin{eqnarray}
a(x,y,z)=a_{0}(x,y,z_{0})Z(z_{0},z).\label{shape}
\end{eqnarray}
%%%%%%%%%%%%%%%%%%%%%%%%%%%%%%%
If the property (\ref{shape}) is required only for a pair of planes,
it can be realized applying the imaging system. Under certain
conditions the shape-invariant property can also be achieved in
the free-space propagation of the beam behind the plane $z=z_{0}$.
The known
examples are the nondiffracting propagation, the Talbot effect and
the self-imaging. For the ideal nondiffracting beam the amplitude
profile $a_{0}$ remains propagation invariant. In the cases of the
Talbot effect and the self-imaging the initial profile reappears
periodically so that (\ref{shape}) is fulfilled for $z=z_{0}+mL$,
where $L$ is the longitudinal period. The longitudinal
periodicity of the Talbot effect requires the lateral periodicity
whose period depends on $L$. In the case of the self-imaging
the amplitude profiles $a_{0}$
can be nonperiodical but they cannot
be chosen arbitrarily. The fundamental difference of the
self-reconstruction effect in comparison with the self-imaging
consists in the fact that the amplitude profile $a_{0}$ to be
periodically reconstructed can be predetermined.
Taking into account that property the
self-reconstruction can be comprehended as the
effect by which the field with the predetermined transverse
amplitude profile $a_{0}$
can be converted into the field described by the complex amplitude $a_{s}$
possessing the following properties:
%%%%%%%%%%%%%%%%
\begin{description}
\item{(a) the complex amplitude $a_{s}$ is the exact solution to
the Helmholtz equation,}
\item{(b) in the free-space propagation the transverse profile of
the field $a_{s}$ reappears periodically with the longitudinal
period $L$, $a_{s}(x,y,z)=a_{s}(x,y,z+mL)$, $m=0,1,2,\cdots$, }
\item{(c) at the planes of reconstruction $z=z_{0}+mL$ the
complex amplitude $a_{s}$ approximates the predetermined
transverse amplitude profile $a_{0}$, $a_{s}(x,y,z_{0}+mL)
\approx a_{0}(x,y,z_{0})$.}
\end{description}
%%%%%%%%%%%%%%%%
The theoretical description of the self-reconstruction was
proposed in \cite{bouchal5}, \cite{horak} and \cite{k+b}.
In practise, the transformation of the signal field with the
predetermined amplitude profile $a_{0}$ into the field exhibiting
the self-reconstruction ability can be performed applying the
spatial filtering in the 4-f optical system. The used spatial
filter is the amplitude mask consisting a set of concentric
annular rings. After spatial filtering the initial field is
represented by the discrete superposition of the nondiffracting
modes propagating with the angular wavenumbers fulfilling
condition of the self-imaging (\ref{sic}). Experimental
verification of the effect was presented in \cite{wagner}.
%%%%%%%%%%%%%%%%%%%%%%%%%%%%%%%
%\section{Controlled 3D spatial localization}
\section{CONTROLLED 3D LIGHT BENDING}
~

Recently, the concept of the spatial shaping of the nondiffracting
fields  and the self-imaging effect have been adopted to realize
the controlled 3D light bending \cite{boukyvn}.
By that method the light can be confined in the volume elements
whose transverse and longitudinal dimensions are comparable to
the wavelength. The transverse intensity profile of the light field
is created as a coherent superposition of the nondiffracting spots whose
position, size and amplitude profile can be controlled.
By the spatial shaping of the nondiffracting fields the single
nondiffracting spots are centred at the required positions.
By the 3D light bending many nondiffracting spots
possessing different angular wavenumbers contribute at the same
position of the transverse plane. As the spots are superposed
constructively at their centra, the resulting spot is strongly
peaked in comparison with the single one. By that way the
predetermined transverse amplitude profile can be  shaped with a
high resolution. As the total field is composed of the
nondiffracting modes with different angular wavenumbers, it is
not nondiffracting. If the angular wavenumbers are conveniently
coupled, the self-imaging effect is achieved. The required
transverse amplitude profiles then appear periodically along the
propagation direction with the controllable period. If the number
of contributing modes with different angular wavenumbers is sufficiently
large, the field amplitude is strongly peaked also along the
direction of propagation. The intensity maxima then appear
periodically at the planes where the required transverse
amplitude profiles are formed. By that way, the controllable
3D light distribution can be produced. In Fig. 18, the comparison
of the spatial shaping of the nondiffracting field and the 3D
light bending is presented. The former case is illustrated in
Fig. 18a.  The required intensity profile is an array of 9 point
sources. That profile is replicated in the
nondiffracting field in such a way that each source point of the  array
excites the nondiffracting spot placed at the position depending
on the position of the corresponding source. The obtained profile
remains invariant under free-space propagation. In Fig. 18b, the 3D light
bending is shown.
In that case, the required profile is created in such a way that
each point source of the array excites many nondiffracting modes
localized at the same position of the transverse plane. As they
are superposed constructively, the resulting spot is sharpened
in comparison with the single mode spot so that the required
profile is replicated with a very good fidelity. Due to the
self-imaging effect, the transverse amplitude profile is available
only at the planes placed periodically along the propagation
direction. Between those planes it disappears due to the
destructive interference of the nondiffracting modes.
Experimental implementation of both the spatial shaping of
nondiffracting fields and of the 3D light bending is proposed in
\cite{bouchol,boukyvn}. Applications of the 3D light bending can
be expected in the design of adaptable optical tweezers enabling
3D manipulation of electrically neutral particles and atoms.
%%%%%%%%%%%%%%%%%%%%%%%%%%%%%%%
\section{PSEUDO-NONDIFFRACTING BEAMS}
~

The nondiffracting beams indicate that the diffraction effects
can be overcome if the propagation of the source-free
monochromatic field described by the homogeneous Helmholtz
equation is considered. However, the ideal nondiffracting beams
carry an infinite energy and their transverse intensity profile
remains unchanged from $-\infty$ to $+\infty$. This is a reason
why the nondiffracting beams cannot be exactly realized. In
experiments only their approximations known as the nearly
nondiffracting or the pseudo-nondiffracting beams can be
obtained. They possess the finite energy and their propagation
properties can be approximated by the properties of the
nondiffracting beams transmitted through the aperture of finite
dimensions or through the Gaussian aperture \cite{overfelt}. The
simplest type of the pseudo-nondiffracting beam known as the
Bessel-Gauss beam \cite{gori} can be obtained directly from the paraxial
form of the Helmholtz equation.
%%%%%%%%%%%%%%%%%%%%%%%%%%%%%%%%
\subsection{Bessel-Gauss beam}
~

The pseudo-nondiffracting Bessel-Gauss beam represents the
spatially modulated Gaussian beam. It can be searched as the
exact solution to the paraxial Helmholtz equation whose
rotationally symmetrical complex
amplitude is assumed as
%%%%%%%%%%%%%%%%%%%%%%%%%%%%%%%%%%%%%
\begin{eqnarray}
a(r,z)=R[g(z)r^{2}]\exp[iZ(z)]a_{g}(r,z),\label{predpbesgaus}
\end{eqnarray}
%%%%%%%%%%%%%%%%%%%%%%%%%%%%%%%%%%%%%
where $a_{g}$ is the complex amplitude of the conventional
Gaussian beam and $R$, $g$ and $Z$ are in the meantime unknown
functions. Substituting (\ref{predpbesgaus}) into the paraxial
Helmholtz equation and applying the denotation
$t^{2}=gr^{2}$  we can write
%%%%%%%%%%%%%%%%%%%%%%%%%%%%%%%%%%%%%%%%
\begin{eqnarray}
t^{2}\frac{d^{2}R}{dt^{2}}+t\frac{dR}{dt}
\left[1-2ik\frac{t^{2}}{2g^{2}}\frac{dg}{dz}
-2\frac{t^{2}}{g}
\left(\frac{2}{w^{2}}+i\frac{k}{R_{g}}\right)\right]
+t^{2}\frac{2k}{g}\frac{dZ}{dz}R=0,\label{prubbes}
\end{eqnarray}
%%%%%%%%%%%%%%%%%%%%%%%%
where $w$ is the bandwidth of the Gaussian beam and $R_{g}$
denotes the radius of the curvature of its wavefront.
If the functions $g$ and $Z$ fulfil equations
%%%%%%%%%%%%%%%%%%%%%%%%%%%%%%%%%%%%%%%%
\begin{eqnarray}
i\frac{k}{2g}\frac{dg}{dz}+\frac{2}{w^{2}}
+i\frac{k}{R_{g}}&=&0,\label{rovprog}\\
\frac{2k}{g}&=&1,\label{rovproz}
\end{eqnarray}
%%%%%%%%%%%%%%%%%%%%%%%%%%%%%%%%%%%%%%%%%%%
then (\ref{prubbes}) represents the special form of the Bessel
equation. In the examined case, its solution is the function
$R$ given by
the Bessel function of the first kind and zero order
%%%%%%%%%%%%%%%%%%%%%%%%%%%%%%%%%%%%%%%%%%%
\begin{eqnarray}
R(t)=J_{0}(t).
\end{eqnarray}
%%%%%%%%%%%%%%%%%%%%%%%%%%%%%%%%%%%%%%
Integrating (\ref{rovprog}) we obtain
%%%%%%%%%%%%%%%%%%%%%%%%%%%
\begin{eqnarray}
ln|g|+C=i2\arctan(z/\overline{q}_{0})-ln|z^{2}+\overline{q}^{2}_{0}|,
\end{eqnarray}
%%%%%%%%%%%%%%%%%%%%%%%%%%%%
where $\overline {q_{0}}$ is the Rayleigh distance of the
Gaussian beam.
If the integration constant $C$ is rewritten by means of
the new constant $K$
%%%%%%%%%%%%%%%%%%%%%%%%%%%
\begin{eqnarray}
C=-2ln|K\overline{q}_{0}|,
\end{eqnarray}
%%%%%%%%%%%%%%%%%%%%%%%%%%%
then the searched function $g$ can be expressed as
%%%%%%%%%%%%%%%%%%%%%%%%%%
\begin{eqnarray}
g=\frac{K^{2}}{\left(1-iz/\overline{q}_{0}\right)^{2}}.
\end{eqnarray}
%%%%%%%%%%%%%%%%%%%%%%%%%%%%%%%%%
If we substitute $g$ into (\ref{rovproz}) the function $Z$
is obtained after integration. Applying it the complex amplitude
of the Bessel-Gauss beam can be rewritten in the form
%%%%%%%%%%%%%%%%%%%%%
\begin{eqnarray}
a(r,z)=J_{0}\left(\frac{K r}{1-iz/\overline{q}_{0}}\right)
\exp\left[i\frac{K^{2}\overline{q}^{2}_{0}}{4k}
\left(\frac{2}{kw^{2}}+i\frac{1}{R_{g}}\right)
(z-i\overline{q}_{0})\right]a_{g}(r,z),
\end{eqnarray}
%%%%%%%%%%%%%%%%%%%%%
where $a_{g}$ is the complex amplitude of the Gaussian beam.
%%%%%%%%%%%%%%%%%%%%%%%%%%%%%%%%%%%%%%%%%%%%%%%
\subsection{Comparison of pseudo-nondiffracting and ideal
nondiffracting beams}
~

The nondiffracting beams possess the sharp  $\delta$-like
angular spectrum represented by a circle at the
Fourier plane. The basic parameter of the spectrum is the single radial
frequency which can be related to the radius of that circle. The radial
frequency is inversely proportional to the transverse dimension
of the intensity profile of the beam. Because the ideal
nondiffracting beams transfer an infinite energy,
they can be realized only approximately in experiments.
They are usually called the pseudo-nondiffracting beams.
The simplest model of the pseudo-nondiffracting beam
can be obtained if the nondiffracting beam is transmitted
through the aperture whose transparency is described by the
Gaussian function. The energy of the transmitted beam
is then finite but its
angular spectrum possesses the spread caused by the aperture.
As a result, the propagation invariance of the transverse
intensity profile is lost but the fundamental differences between
propagation properties of the conventional and the
pseudo-nondiffracting beams
still exist.
In \cite{bouchal3} they were demonstrated on the
composition of the angular spectra of both types of beams.
The main results of that analysis can be concluded as follows:
%%%%%%%%%%%%%%%%%%%%%%%
\begin{itemize}
\item{The conventional beam with the spatial bandwidth
$2\Delta r$ transmitted through the aperture with the transverse
dimension $2\Delta R$
possesses the angular spectrum with the uncertainty
$2\Delta\theta$. If $\Delta R>>\Delta r$, the spread of the
angular spectrum depends on the transverse size $2\Delta r$ of the
beam impinging on the aperture. This dependence can be written as
the uncertainty relation
\begin{eqnarray}
\Delta r\Delta\theta=const.
\end{eqnarray}
}

\item{The pseudo-nondiffracting beam is obtained if the nondiffracting
beam with the intensity spot size $2\Delta r$ is transmitted through the
aperture with the transverse dimension $2\Delta R$. If $\Delta R>>\Delta r$,
the spread of the angular spectrum of the pseudo-nondiffracting
beam is given by the relation
\begin{eqnarray}
\Delta R\Delta\theta=const.\label{unrel}
\end{eqnarray}
The spread
of the angular spectrum of the pseudo-nondiffracting
beam depends only on $\Delta R$.
If the size of the aperture $2\Delta R$ is constant,
$\Delta\theta$ remains unchanged even if the spot size $2\Delta
r$ of the beam impinging on the aperture decreases. This
property represents the fundamental difference between the
conventional and the pseudo-nondffracting beams.
It represents an essence of the
pseudo-nondiffracting propagation.
In this sense we can speak of the diffraction elimination.}
The property (\ref{unrel}) is graphically illustrated in Fig. 19.
The transverse intensity profiles of the nondiffracting beams
impinging on the Gaussian aperture are illustrated in Fig. 19a and
19c. The corresponding annular spectra are illustrated in Fig.
19b and 19d. As is obvious, the reduction of the size of the
intensity spot of the input beam changes only diameter of the
annular ring but its width important for the diffractive
divergence of the beam remains unchanged.\\
~

For the constant intensity spot size of the impinging nondiffracting beam
$2\Delta r$, the spread of the angular spectrum $2\Delta\theta$
and also the  diffractive divergence of the transmitted
pseudo-nondiffracting beam
$2\Delta\vartheta$ can be reduced if the window is enlarged. As the
impinging nondiffracting beam falls to zero in the transverse
direction very slowly, the reduction of the diffractive spread
is associated with the increase of the energy consumption.

\item{The price payed for the reduction of the
diffraction effects is shortening of the longitudinal range of
the beam existence $\Delta z$. It is given by the
relation
\begin{eqnarray}
\Delta z\Delta\theta=const\Delta r.
\end{eqnarray}
For the constant size of the aperture $2\Delta R$, the longitudinal
range $\Delta z$ is shortened if the spot size of the beam
impinging on the window decreases.}
\end{itemize}
%%%%%%%%%%%%%%%%%%%%%%%%%%%%%%%%%%%%%%%%%%%%%%%%%%%%%%%%%%%%%
\section{EXPERIMENTS AND APPLICATIONS}
~

The monochromatic coherent nondiffracting beam can be
comprehended as an interference field produced by the
superposition of plane waves whose propagation vectors form the conical
surface. Such interference field can be realized in a good
approximation by several methods
(see \cite{lapointe} for a review). The simple way how to
generate the pseudo-nondiffracting beam  whose transverse
amplitude profile resembles the zero-order Bessel function
of the first kind $J_{0}$ was proposed by Durnin et al
\cite{durninetal}.
The used experimental setup is illustrated in
Fig. 20.
In that experiment
the spatially filtered and expanded laser beam
illuminates the annular ring mask placed at the front focal plane of
the lens. The mask serves as a secondary source generating
spherical waves. Each of them is transformed by the lens to the
beam of parallel rays propagating with the angle $\theta_{0}$
with respect to the optical axis.
The beam-like interference field with the $J_{0}$ transverse
amplitude profile appears in the interference region behind the lens
where the beams intersect.
The spot size of the generated beam depends on the angle
$\theta_{0}$. The radius of the beam central spot $r_{0}$ can be
approximated by $r_{0}\approx \lambda/sin{\theta_{0}}$. For
angles close to $\pi/2$ the light tubes with the size comparable
to the wavelength can be obtained. The longitudinal distance $L$
where the beam is available without changes of its transverse
intensity profile depends on the angle $\theta_{0}$ and on the
size of the aperture of the Fourier lens.
The distance where the beam of the given spot size propagates
without apparent changes of its intensity profiles can be
controlled by the size of the lens aperture.
The pseudo-nondiffracting beam can be
obtained even if an arbitrary amplitude and/or phase modulation
is applied at the plane of the annular mask. For example,
the Bessel-like beams of higher-order can be realized if the
spiral phase plate is adjacent to the annular ring.
Production of such phase plate is a complicated technical
problem. It can be successfully prepared applying the
photolitographic techniques.
If the phase plate causes the azimuthal phase modulation given by
$\exp(im\varphi)$, the $m$-th order Bessel beam is approximately
generated. Its axial intensity is equal to zero so that the beam
is dark and exhibits phase properties of the optical vortices.
The pseudo-nondiffracting beams can also be realized by the
azimuthal amplitude modulation of the annular ring mask.
For example, it can be achieved if the annular mask is
illuminated by the one-dimensional strip pattern with the
Gaussian profile. The obtained beam represents a good
approximation of the Mathieu beam. The pseudo-nondiffracting
beam can also be obtained as an interference field produced by the
discrete superposition of plane waves. In that case the annular
ring mask is transparent only at the finite number of points.
It can be simply realized by means of the auxiliary amplitude
mask. If the annular ring mask is illuminated by the
source whose correlation properties can be controlled the
partially-coherent nondiffracting beams can be generated.
In \cite{bouchal6} their properties were examined for the illumination
realized applying the Gaussian Shell-model source.
The self-imaging effect can be experimentally realized if the
amplitude mask consisting of a set of annular rings with the
required diameters is used. The experimental verification of that
effect was presented in \cite{bouchal4}. Another possible way how
to obtain an interference field of plane waves whose wave
vectors form the conical surface is based on the use of the
refractive axicon illuminated by the collimate laser beam
\cite{herman,scott}.
An advantage of that method is the high efficiency with which the
power of the common laser beam
can be converted into the pseudo-nondiffracting form.
In \cite{tur,vas} it was shown that the action of the annular
mask or the refractive axicon can be alternatively performed
by the computer-generated axicon-type hologram. In that way
the beam approximating the zero-order Bessel beam was obtained
with the relative high conversion efficiency approaching
50\% \cite{lee1}.
The dark higher-order Bessel beams were also successfully
generated by the holographic means \cite{lee2,paterson}.
The simple but efficient method providing a good approximation
of the zero-order Bessel beam can be realized applying
the centrally obscured lens
exhibiting the spherical aberration \cite{herman}.
In \cite{karim} a possibility to convert the Gaussian beam to the
zero-order Bessel beam applying the two-element refracting system
was examined. In that method the conversion efficiency is
good but the system is hardly realizable because the optical
elements with the aspheric surfaces are required.
The general pseudo-nondiffracting patterns useful for the
optical interconnection applications were realized by means of the
magneto-optic spatial light modulators \cite{davis1,davis}.
The pseudo-nondiffracting beams can be also generated directly
at the laser resonator. The special resonator construction was
proposed in \cite{uehara}. Recently,  the experiment enabling
generation of the nondiffracting beams with the controllable
spatial coherence was proposed and realized \cite{bouchal6}.
It is based on the use of the pseudothermal , so called
Gaussian Shell-model source. In that case the optical scheme in Fig. 20
remains unchanged but the coherent laser beam illuminating the
annular ring is replaced by the source shown in Fig. 21.
The coherent laser beam is focused to the rotating diffuser where
its phase is randomized. The beam spot created at the diffuser
serves as a spatially incoherent source illuminating the annular
ring. During free propagation between the diffuser and the annular
mask the spatial coherence of the beam is increased so that the
annular ring is illuminated by the partially coherent light.
Its coherence properties can be continuously changed by the
change of the beam spot at the rotating diffuser and described
mathematically applying the Van Cittert-Zernike theorem.
The light illumination can be changed from fully incoherent to
nearly fully coherent.
The change of the spatial coherence was applied to the optical
set-up enabling generation of the nondiffracting beams with the
predetermined transverse intensity profile \cite{bouchol}.
The optical scheme is shown in Fig. 22. The required amplitude
profile of the beam is predicted by the source array. The mask is
then illuminated by the light of the controllable spatial
coherence. The required amplitude profile is then replicated as
the nondiffracting pattern. The method works in such a way that
each point of the source array excites the nondiffracting spot
whose size and form can be driven by the experiment geometry.
The position of the nondiffracting spot is defined by the
position of the source point. The nondiffracting pattern is
obtained as a superposition of the nondiffracting spots.
The fidelity of the replication is highest if the nondiffracting
modes are superposed incoherently.

The unique properties of the nondiffracting beams are useful for
both the technical and physical applications.
The propagation invariance of their transverse intensity profile
is applicable in metrology for scanning optical systems
\cite{arimoto}.  The nondiffracting beams are also suitable for
large-scale straitness measurement and other large size
measurement \cite{wang} because are much less influenced
by atmospheric turbulence than other beams \cite{aruga}.
Attention was concentrated also to the imaging applications of the
nondiffracting beams. It was verified that the imaging
realized with nondiffracting beams can provide an extremely long
focal dept.  In \cite{wing}, a kilometer-long imaging was
proposed and examined. An increasing attention is devoted to the
nondiffracting beams for their applicability in nonlinear optics.
In \cite{wulle}, it was shown that the nondiffracting Bessel beam
can be viewed as a light beam with the tunable wavelength.
Due to that property,
the phase-matched second-harmonic generation at angles
usually not suited for phase matching in a KDP crystal was
performed. The application of the nondiffracting beams to the
third-harmonic generation \cite{tewari} and to the \v{C}erenkov
second-harmonic generation in bulk optical crystals was also
proposed \cite{pandit}. An efficient conical emission of light in
Raman scattering stimulated by nondiffracting Bessel beam was
examined in \cite{niggl}. The Bessel pump beam was applied also
to the design of the distributed-feedback laser \cite{klewitz}.
The nondiffracting beams were applied to increase the sensitivity
of the measurement of the nonlinear refractive index by the
$Z$-scan method \cite{hughes}.
The radially polarized nondiffracting beams possess the strong
longitudinal component of the electric field. That component can
accelerate the particles of the electron beam propagating nearly
collinearly with the laser nondiffracting beam \cite{tidwell}.
Recently, the application of the nondiffracting beams
significantly improved the techniques for manipulations of
micrometer-sized particles. The self-reconstruction ability of
the nondiffracting beam enables to manipulate ensembles of
particles simultaneously in multiple planes \cite{hegner,dholakia}.
The nondiffracting vortex beams are perspective for the research
focused to the transfer of the orbital angular momentum to the
particles. The obtained results are promising for realization of
the light motors whose rotors can be forced by the laser beam.
The nondiffracting beams are also perspective for the atom
guiding in their optical potential \cite{arlt}.
%%%%%%%%%%%%%%%%%%%%%%%%%%%%%%%%%%%%%%%%
\section{CONCLUSIONS}
~

In the paper, theoretical concepts, mathematical methods of
description and numerical simulations of the nondiffracting
propagation were reviewed. The particular attention was focused
to the physical properties of nondifracting beams and to their
physical applications such as 3D light synthesis, self-imaging
and self-healing effects and the transfer of the orbital angular
momentum to the material particles. The experimental realization
of nondiffracting beams and their technical aplications were also
presented.
%%%%%%%%%%%%%%%%%%%%%%%%%%%%%%%%%%%%%%
\section*{Acknowledgements}
This research was supported by the projects LN00A015 and CEZ J14/98
of the Ministry of Education of the Czech Republic and by EU
grant under QIPC project IST-1999-13071 (QUICOV).
Dr. J. Wagner and prof. J. Pe\v{r}ina are acknowledged for experimental
verifications of the theoretical results and for lasting
interest, furtherance and inspiration.
%%%%%%%%%%%%%%%%%%%%%%%%%%%%%%%

%%%%%%%%%%%%%%%%%%%%%%%%%%%%%%%%
\section*{Figure captions}
%%%%%%%%%%%%%
Fig. 1\\
The Bessel beam  as a superposition of the travelling waves
given by the Hankel functions. The outgoing wave travelling away
from the axis (a) and the incoming wave travelling towards the
axis (b) create the standing wave represented by the Bessel
function (c).\\[5pt]
Fig. 2\\
The transverse intensity profiles of the bright zero-order Bessel beam
(a) and the dark first-order Bessel beam (b).\\[5pt]
Fig. 3\\
The helical wavefronts of the nondiffracting vortex beams for the
topological charge (a) $m=1$ and (b) m=2.\\[5pt]
Fig. 4\\
Visualisation of the optical vortices. Interference of the
optical vortex $m=1$ with the spherical wave results in spiral
and fork-like interference patterns (a) and (b).
Interference of the vortex with the topological charge $m=2$
is in (c) and (d).\\[5pt]
Fig. 5\\
The transverse intensity profile of the Mathieu beams for
parameters (a) $w_{0}=4\nu_{0}$ and (b) $w_{0}=2\nu_{0}$.\\[5pt]
Fig. 6\\
The caleidoscopic nondiffracting patterns obtained as a
discrete superposition of the finite number of plane waves:
(a) $N=5$, (b) $N=10$, (c) $N=15$, and (d) $N=30$.\\[5pt]
Fig. 7\\
Illustration of the nondiffracting field as a superposition of
five cosine gratings with the same period and various
orientations.\\[5pt]
Fig. 8\\
Nondiffracting beams with the variable spatial coherence.
The fully coherent dark vortex beam (a) is continuously
changed to the bright nondiffracting beam (b)-(d) if the spatial
coherence is decreased.\\[5pt]
Fig. 9\\
Illustration of the degree of angular correlation of the plane
waves creating the partially coherent nondiffracting beam.\\[5pt]
Fig. 10\\
Change of the vortex topology caused by the change of the
spatial coherence. The vortex of the coherent beam (a) vanishes
if the spatial coherence is decreased (b)-(d).\\[5pt]
Fig. 11\\
Spatial shaping of coherent and incoherent nondiffracting fields.
The required intensity profile (a) is replicated in the
nondiffracting beam created with incoherent light (b) and coherent
light (c). In (d) and (c) the parameter $\alpha$ of the nondiffracting
spots is increased so that their size is reduced.\\[5pt]
Fig. 12\\
The same as in Fig. 11 but for the continuous required intensity
profile.\\[5pt]
Fig. 13\\
Illustration of the vectorial electromagnetic nondiffracting beams.
The azimuthal polarization of the electric field
(a) and the radially polarized magnetic field (b). The short
arrows illustrate the magnitude and direction of the transverse
components of the field vectors at the separate points of the
transverse plane.\\[5pt]
Fig. 14\\
Experimental verification  of the resistance of the nondiffracting
beam against amplitude and phase perturbations. The nondiffracting beam
impinging on the nontransparent obstacle is fully revived during
free propagation behind the obstacle.\\[5pt]
Fig. 15\\
Healing of the spatial distribution of the orbital angular
momentum of the disturbed nondiffracting vortex beam.
The initial beam with the spatial distribution of the orbital
angular momentum (a) interacts with the complex object which
takes the orbital angular momentum and rotates. After
interaction the spatial distribution of the beam orbital angular
momentum is disturbed (b)-(c) but during free propagation is
revived to the nearly initial form (d).\\[5pt]
Fig. 16\\
Healing of the vortex topology after interaction with the complex
object accompanied by
the exchange of the orbital angular momentum.\\[5pt]
Fig. 17\\
The self-imaging effect obtained due to the coherent
superposition of the nondiffracting modes. The transverse
intensity profile appears periodically along the propagation
direction $z$ with the longitudinal period controllable by the
choice of the angular wavenumbers of the nondiffracting
modes.\\[5pt]
Fig. 18\\
Comparison of the controlable shaping of the nondiffracting
fields (a) with the 3D light bending (b). By the light bending
the required transverse profile appear only at the near vicinity
of the planes placed periodically along the propagation direction
so that the light is confined at 3 dimensions. By that method the
light can be localized in the volume elements with dimensions
comparable to the wavelength.\\[5pt]
Fig. 19\\
Illustration of spatial spectrum of the pseudo-nondiffracting
beam.
The pseudo-nondiffracting beam with intensity profile (a)
possesses the spatial spectrum of the form (b). If the intensity
profile is rescaled (c),  the spread of the spatial spectrum
(width of the ring) remains unchanged (d).
On that behavior the unique properties of the
pseudo-nondiffracting beams are based and it cannot be achieved by the
conventional beams.\\[5pt]
Fig. 20\\
Experimantal set-up enabling generation of the
pseudo-nondiffracting beam.\\[5pt]
Fig. 21\\
Gaussian Shall-model source used for generation of the partially
coherent pseudo-nondiffracting beams.\\[5pt]
Fig. 22\\
Optical set-up for generation of the nondiffracting fields with
the predetermined intensity profile.
%%%%%%%%%%%%%%%%%%%%%%%%%%%%%%%%
%\newpage
%\includegraphics[width=15cm]{fig1}
\end{document}